\documentclass{article}
\begin{document}
\newcommand{\be}{\begin{eqnarray}}
\newcommand{\ee}{\end{eqnarray}}
\newcommand{\ra}{\rightarrow}
\newcommand{\la}{\leftarrow}
\newcommand{\noi}{\noindent}
\title{Quasiprobability Based Criterion for Classicality
and Separability of States of Spin-1/2 Particles
\footnote{Phys.Rev. A{\bf 86}, 052111 (2012)}}
\author{R.R.Puri\footnote{e-mail: ravirpuri@gmail.com}\\
Training School,
Bhabha Atomic Research Centre\\
and\\
Homi Bhabha National Institute\\
Anushaktinagar, Mumbai-400094, India.}
\date{}
\maketitle
\noi
\begin{abstract}
\noindent
A sufficient condition for a quantum state of a system of spin-1/2
particles (spin-1/2s) to admit a local hidden variable (LHV)
description i.e. to be classical is the separability of the density
matrix characterizing its state, but not all classical states are
separable. This leads one to infer that separability and classicality
are two different concepts. These concepts are examined here in
the framework of a criterion for identifying classicality of a system
of spin-1/2s based on the concept of joint quasiprobability (JQP) for
the eigevalues of spin components (R.R.Puri,J.Phys.{\bf A29}, 5719
(1996)). The said criterion identifies a state as classical if a
suitably defined JQP of the eigenvalues of spin components in suitably
chosen three orthogonal directions is non-negative. In agreement with
other approaches, the JQP based criterion also leads to the result
that all non-factorizable pure states of two spin-1/2s are
non-classical. In this paper it is shown that the application of the
said criterion to mixed states suggests that the states it identifies
as classical are also separable and that there exist states which,
identified as classical by other methods, may not be identified as
classical by the criterion as it stands. However, the results in
agreement with the known ones are obtained if the criterion is
modified to identify also those states as classical for which the JQP
of the eigenvalues of the spin components in two of the three
prescribed orthogonal directions is non-negative. The validity of the
modified criterion is confirmed by comparing its predictions with
those arrived at by other methods when applied to several mixed states
of two spin-1/2s and the Werner like state of three spin-1/2s
(G.Toth and A.Acin,Phys.Rev. {\bf A74}, 030306(R) (2006)). The JQP
based approach, formulated as it is along the lines of the
$P$-function approach for identifying classical states of the
electromagnetic field, offers a unified approach for systems of
arbitrary number of spin-1/2s and the possibility of linking
classicality with the nature of the measurement process.
\end{abstract}

\section{Introduction}

In view of its importance in quantum computing and quantum information
processing~\cite{chuang}, identification of exclusively quantum
correlations in a collection of identical two level systems, hereafter
referred to as a system of spin-1/2 particles (spin-1/2s), continues to
attract a great deal of attention~\cite{guhne,horormp}. The
correlations are classical if they can be described in terms of a
purely classical local hidden variable (LHV) theory. A sufficient
condition for a quantum state of a system consisting of two subsystems
to admit a LHV description is that the density matrix $\hat\rho$
characterizing its state be separable i.e. it be expressible
as~\cite{werner}
\be
\hat\rho=\sum_{i}p_i\hat\rho^{(1)}_i\otimes\hat\rho^{(2)}_i,
\quad 0\le p_i\le 1, \quad \sum_{i}p_i=1,
\label{n1}
\ee
where $\hat\rho^{(1)}_i$ and $\hat\rho^{(2)}_i$ are the density
matrices of the subsystems numbered 1 and 2. A necessary condition
for separability defined in (\ref{n1}) was derived by
Peres~\cite{peres}. It was subsequently shown that the Peres
condition is also sufficient if the Hilbert space of each of the two
subsystems is two dimensional or if the dimension of the Hilbert space
of one of the systems is two and that of the other is
three~\cite{horodecki1}. Thus, for a system of two spin-1/2s, the
Peres criterion constitutes necessary and sufficient condition for
separability.

Starting with the work of Werner~\cite{werner}, the relationship
between entanglement and non-locality has been investigated
extensively (see for example references~\cite{barrett}-\cite{masanes}
and the references therein). In particular, by constructing a
non-separable mixed state of two subsytems which nevertheless admits
LHV description under projective measurements, Werner~\cite{werner}
showed that separability is a sufficient but not a necessary condition
for classicality. This result has since been generalized to general
measurement processes~\cite{barrett} and to tripartite
systems~\cite{toth}.

In this paper we formulate an approach to identifying classicality
in a system of spin-1/2s along the lines of that followed for
classifying the states of the electromagnetic (e.m.) field as
classical and non-classical which is based on the concept of
quasiprobability distribution (QPD) function corresponding to a
quantum state. Recall that a QPD is constructed by considering
the quantum observables as classical random variables whose
probability distribution is expressed in terms of averages of products
of observables. The observables may correspond to non-commuting
operators. A QPD is constructed by replacing the averages of products
of observables by the expectation values of the corresponding
operators by choosing the order in which the  non-commuting operators
be placed in the products of corresponding observables. There is thus
a QPD for each ordering of product of non-commuting operators. The
classicality of a state may be identified as that of the QPD in a
specified operator ordering. The specification of the ordering for the
said purpose is based on physics considerations like the relation
between the operator ordering and the measurement process or some
other desirable physics aspect. For example, the classicality of the
states of the electromagnetic (e.m.) field is defined in terms of the
QPD called Sudarshan-Glauber P-function. That function is the QPD
corresponding to the ordering of the field creation and annihilation
operators appropriate for those processes of measurement, like the
ones by photodetectors, which measure averages of normally ordered
operators~\cite{cahil, puri}.   

The physics consideration for constructing the QPD relevant to
identifying quantum nature of correlations in a system of spin-1/2s 
could be that it should identify every state of a spin-1/2 as also
every uncorrelated state of a collection of spin-1/2s as classical.
The so called phase space QPDs for spin-1/2s, constructed in analogy
with those for the e.m. field, do not fulfill the desired conditions
~\cite{puri}-\cite{giraud}. The desired end is achieved by invoking
the concept of joint quasiprobability (JQP)~\cite{puri},
\cite{feynman}-\cite{puri1} for the distribution of the eigenvalues of
the components of spins in various directions. Based on that concept,
and by demanding that any state of a single spin-1/2 and any product
of single spin-1/2 states be classical, a criterion has been
introduced in Ref.~\cite{puri1} to classify the
states as classical and non-classical. The said criterion identifies a
state as classical if a suitably defined JQP of the eigenvalues of
spin components in suitably chosen three orthogonal directions is
non-negative i.e. classical. It leads to the conclusion that any
non-factorizable pure state of two spin-1/2s is non-classical. This
result is arrived at also by other approaches~\cite{gisin}.

When applied to mixed states, it is found that the said criterion
as it stands may leave out from the set of states it identifies as
classical some separable states. It also does not identify those
non-separable states as classical which are so identified by other
methods. In this paper we propose a modified JQP based criterion which
is free of the abovementioned lacunae. The modified criterion
identifies as classical, not only those states whose JQP for
suitably chosen three orthogonal components of each spin is
non-negative, but also those for which the JQP for any two of those
three orthogonal components is non-negative. The validity of the
modified criterion is confirmed by examining the classicality and
separability of some of those mixed states of two spin-1/2s and a
state of three spin-1/2s whose said properties are known following
other approaches.

The paper is organized as follows. The Sec.\ref{sec2} summarizes the
main results of the theory of JQP for a system of spin-1/2s as
formulated in~\cite{puri2,puri1} and states the JQP based criterion of
classicality as it stands and its proposed modified version. The
conditions of classicality of certain mixed states of two spin-1/2s
and that of a system of three spin-1/2s are derived using the modified
JQP based criterion and the results are compared with those found
by other methods. The Sec.\ref{sec4} summarizes the conclusions.

\section{Joint Quasiprobability for System of Spin-1/2s}~\label{sec2}

In local hidden variable (LHV) theory a spin-1/2 is visualized as a
vector ${\bf S}$ whose components along any direction can assume
two values, say $\pm 1/2$, and is assumed to be under the influence of
some unknown "hidden" causes or variables acting randomly. The random
influence of the hidden variables results in the components of the spin
in any direction acquiring randomly the values 1/2 and -1/2. The
properties of the spin may then be described in terms of the
probability distribution functions $f(S_{a_1},S_{a_2},
\ldots,S_{a_N)})$ for the components of the spin in various directions
where
\be
S_{a_i}={\bf S}\cdot{\bf a}_i,
\label{new2}
\ee
is the component of spin in the direction ${\bf a}_i$.
Now, let $p(\epsilon_{a_1},\epsilon_{a_2},\ldots,\epsilon_{a_N})$
($\epsilon_{a_i}=\pm 1$) denote the joint probability for the
components of the spin along the directions ${\bf a}_1,{\bf a}_2,
\ldots,{\bf a}_N$ to have the values $\epsilon_{a_1}/2,\epsilon_{a_2}/2,
\ldots,\epsilon_{a_N}/2$ respectively so that
\be
&&f\left(S_{a_1},S_{a_2},\ldots,S_{a_N}\right)\nonumber\\
&&=\sum_{\epsilon_1,\epsilon_2,
\ldots,\epsilon_N=\pm 1}\left[\delta\left(S_{a_1}-
\frac{\epsilon_{a_1}}{2}\right)
\delta\left(S_{a_2}-\frac{\epsilon_{a_2}}{2}\right)
\cdots\delta\left(S_{a_N}-\frac{\epsilon_{a_N}}{2}\right)\right]
\nonumber\\
&&~~~~~~~~~~~~~~~~~~~~
\times p(\epsilon_{a_1},\epsilon_{a_2},\ldots,\epsilon_{a_N}).
\label{new2m}
\ee
From this it is straightforward to see that~\cite{puri2,puri1}
\be
p(\epsilon_{a_1},\epsilon_{a_2},\ldots,\epsilon_{a_N})
=\Big\langle\Big(\frac{1}{2}+\epsilon_{a_1}S_{a_1}\Big)
\Big(\frac{1}{2}+\epsilon_{a_2}S_{a_2}\Big)
\cdots\Big(\frac{1}{2}+\epsilon_{a_N}S_{a_N}\Big)\Big\rangle,
\label{new1}
\ee
where the angular bracket denotes average with respect to the
distribution function $f\left(S_{a_1},S_{a_2},\ldots,S_{a_N}\right)$:
\be
\langle A\rangle=\int~Af\left(S_{a_1},S_{a_2},\ldots,S_{a_N}\right)
\prod_{i=1}^{N}{\rm d}S_{a_i}.
\label{new2p}
\ee
The joint probability distribution for two spins can be similarly
defined and shown to be expressible as
\be
&&p(\epsilon^{(1)}_{a_1},\epsilon^{(1)}_{a_2},\ldots,
\epsilon^{(1)}_{a_M};\epsilon^{(2)}_{b_1},\epsilon^{(2)}_{b_2},
\ldots,\epsilon^{(2)}_{b_N})\nonumber\\
&&=\Big<\Big(\frac{1}{2}+\epsilon^{(1)}_{a_1}S_{a_1}\Big)
\Big(\frac{1}{2}+\epsilon^{(1)}_{a_2}S_{a_2}\Big)
\cdots\Big(\frac{1}{2}+\epsilon^{(1)}_{a_M}S_{a_M}\Big)
\nonumber\\
&&~~~~~~\Big(\frac{1}{2}+\epsilon^{(2)}_{b_1}S_{b_1}\Big)
\Big(\frac{1}{2}+\epsilon^{(2)}_{b_2}S_{b_2}\Big)
\cdots\Big(\frac{1}{2}+\epsilon^{(2)}_{b_N}S_{b_N}\Big)
\Big>.
\label{new3}
\ee
The function $p(\epsilon^{(1)}_{a_1},\epsilon^{(1)}_{a_2},\ldots,
\epsilon^{(1)}_{a_M};\epsilon^{(2)}_{b_1},\epsilon^{(2)}_{b_2},
\ldots,\epsilon^{(2)}_{b_N})$ stands for the probabilty of finding
the components of spin number 1 to have values
$\epsilon^{(1)}_{a_1}/2,\epsilon^{(1)}_{a_2}/2,\ldots,
\epsilon^{(1)}_{a_M}/2$
in the directions ${\bf a}_1,{\bf a}_2,\ldots,{\bf a}_M$ and
the components of spin number 2 to have the values
$\epsilon^{(2)}_{b_1}/2,\epsilon^{(2)}_{b_2}/2,\ldots,
\epsilon^{(2)}_{b_N}/2$ in the directions
${\bf b}_1,{\bf b}_2,\ldots,{\bf b}_N$ with
$\epsilon^{(1)}_{a_i},\epsilon^{(2)}_{b_i}=\pm 1$.

Now, a spin-1/2 in quantum theory is described by the vector operator
$\hat{\bf S}$ and its state is characterized by a density matrix
$\hat\rho$ using which one can evaluate expectation values of relevant
operators. As we know, there is no concept of joint probability of
assigning definite values for two non-commuting observables and hence
there is no place for the joint probabilities, like the ones defined
in (\ref{new1}) and (\ref{new3}), in the quantum theory.

However, the concept of quasiprobabilities has been found useful to
understand the signatures of quantum behaviour. That concept in the
case of single spin-1/2 may be introduced (i) by replacing the
classical random variables $S_{a}$ by the operator $\hat S_{a}$
which obey the commutation relation
\be
[\hat S_{a},~\hat S_{b}]={\rm i}({\bf a}\times{\bf b})\cdot\hat{\bf S},
\label{comel}
\ee
and the anti-commutation relation
\be
\hat S_a\hat S_b+\hat S_b\hat S_a=\frac{{\bf a}\cdot{\bf b}}{2},
\label{new9}
\ee
(ii) by assigning a rule, called the Chosen Ordering (CO), for
ordering operators in a product of non-commuting operators, and
(iii) by replacing the average therein as the quantum mechanical
expectation value wherein the system is described by a density matrix
$\hat\rho$ and the expectation values of an operator $\hat A$ is given
by ${\rm Tr}(\hat A\hat\rho)$. The expression in quantum theory
corresponding to (\ref{new1}) then reads
\be
&&p(\epsilon_{a_1},\epsilon_{a_2},\ldots,\epsilon_{a_N})\nonumber\\
&&={\rm Tr}\Big[\hat\rho\Big\{\Big(\frac{1}{2}
+\epsilon_{a_1}\hat S_{a_1}\Big)
\Big(\frac{1}{2}+\epsilon_{a_2}\hat S_{a_2}\Big)
\cdots\Big(\frac{1}{2}+\epsilon_{a_N}\hat S_{a_N}\Big)
\Big\}_{{\rm CO}}\Big],
\label{new6}
\ee
where 'CO' stands for Chosen Ordering of the operator product.
In the same way, the quantum analog of the joint quasiprobability for
two spins reads
\be
&&p(\epsilon^{(1)}_{a_1},\epsilon^{(1)}_{a_2},\ldots,
\epsilon^{(1)}_{a_M};\epsilon^{(2)}_{b_1},\epsilon^{(2)}_{b_2},
\ldots,\epsilon^{(2)}_{b_N})\nonumber\\
&&={\rm Tr}\Big[\hat\rho\Big\{\Big(\frac{1}{2}
+\epsilon^{(1)}_{a_1}\hat S_{a_1}\Big)
\Big(\frac{1}{2}+\epsilon^{(1)}_{a_2}\hat S_{a_2}\Big)
\cdots\Big(\frac{1}{2}+\epsilon^{(1)}_{a_M}\hat S_{a_M}\Big)
\nonumber\\
&&~~~~~~\Big(\frac{1}{2}+\epsilon^{(2)}_{b_1}\hat S_{b_1}\Big)
\Big(\frac{1}{2}+\epsilon^{(2)}_{b_2}\hat S_{b_2}\Big)
\cdots\Big(\frac{1}{2}+\epsilon^{(2)}_{b_N}\hat S_{b_N}
\Big)\Big\}_{{\rm CO}}\Big].
\label{new7}
\ee
In what follows, it will be seen that an ordering of interest is the
symmetric ordering in which the c-number product is replaced by the
operator product by following correspondence:
\be
S_aS_b&\rightarrow &\frac{1}{2}[\hat S_a\hat S_b+\hat S_b\hat S_a]
=\frac{{\bf a}\cdot{\bf b}}{4},
\nonumber\\
S_aS_bS_c&\ra &\frac{1}{12}[\hat S_a(\hat S_b\hat S_c
+\hat S_c\hat S_b)+(\hat S_b\hat S_c+\hat S_c\hat S_b)\hat S_a
\nonumber\\
&+&(a\ra b, b\ra c, c\ra a)+(a\ra c, c\ra b, b\ra a)\nonumber\\
&=&\frac{1}{12}[{\bf b}\cdot{\bf c}\hat S_a
+{\bf a}\cdot{\bf c}\hat S_b+{\bf a}\cdot{\bf b}\hat S_c].
\label{new8}
\ee
In writing the equation above we have invoked the anti-commutation
relation (\ref{new9}).

On using the correspondence above as the 'CO', the expression
(\ref{new6}) for the probability distribution of the components
of single spin-1/2 in three mutually orthogonal directions ${\bf a}_1,
{\bf b}_1, {\bf c}_1$ assumes the form
\be
p(\epsilon_{a_1},\epsilon_{b_1},\epsilon_{c_1})
=\frac{1}{2^2}{\rm Tr}\Big[\hat\rho\Big(\frac{1}{2}
+\epsilon_{a_1}\hat S_{a_1}+\epsilon_{b_1}\hat S_{b_1}
+\epsilon_{c_1}\hat S_{c_1}\Big)\Big].
\label{new6m}
\ee
Similarly, the expression (\ref{new7}) for the JQP for the
components of spin number 1 in the orthogonal directions
${\bf a}_1,{\bf b}_1, {\bf c}_1$ and those of spin number 2 in the
orthogonal directions ${\bf a}_2,{\bf b}_2, {\bf c}_2$ in the 'CO'
defined in (\ref{new8}) would read
\be
&&p(\epsilon^{(1)}_{a_1},\epsilon^{(1)}_{b_1},\epsilon^{(1)}_{c_1};
\epsilon^{(2)}_{a_2},\epsilon^{(2)}_{b_2},\epsilon^{(2)}_{c_2})
\nonumber\\
&&=\frac{1}{2^4}{\rm Tr}\prod_{j=1}^{2}\Big[\hat\rho\Big(\frac{1}{2}
+\epsilon^{(j)}_{a_j}\hat S_{a_j}+\epsilon^{(j)}_{b_j}\hat S_{b_j}
+\epsilon^{(j)}_{c_j}\hat S_{c_j}\Big)\Big].
\label{8p}
\ee
The generalization of the considerations above leads to the following
expression for the JQP for the components of three spin-1/2s in the
mutually orthogonal directions directions (${\bf a}_i,{\bf b}_i,
{\bf c}_i)$ ($i=1,2,3$):
\be
&&p(\epsilon^{(1)}_{a_1},\epsilon^{(1)}_{b_1},\epsilon^{(1)}_{c_1};
\epsilon^{(2)}_{a_2},\epsilon^{(2)}_{b_2},\epsilon^{(2)}_{c_2};
\epsilon^{(3)}_{a_3},\epsilon^{(3)}_{b_3},\epsilon^{(3)}_{c_3})
\nonumber\\
&&=\frac{1}{2^6}{\rm Tr}\prod_{j=1}^{3}\Big[\hat\rho\Big(\frac{1}{2}
+\epsilon^{(j)}_{a_j}\hat S_{a_j}+\epsilon^{(j)}_{b_j}\hat S_{b_j}
+\epsilon^{(j)}_{c_j}\hat S_{c_j}\Big)\Big].
\label{8p3c}
\ee
The expressions (\ref{new6m})-(\ref{8p3c}) are central for
classifying the states as classical or non-classical.

Now, recall that the states of the e.m. field are labelled classical
or non-classical on the basis of the classicality or otherwise of the
quasiprobability function, called Sudarshan-Glauber P-function.
The choice of the P-function for the purpose is based on the
fact that it serves as a probability distribution function for the
averages of normally ordered field creation
and annihilation operators and that the process of photo-detection
measures the average of normally ordered field operators. Since in the
case of a system of spins, it is the correlations between the spins
which are the desired measure of the quantum nature, it would be
appropriate to formulate the criterion for identifying the
classicality of the states of a system of spin-1/2s by demanding that
the chosen quasiprobability identifies any state of single spin-1/2 as
also any product of single spin-1/2 states to be classical so that
non-classicality, if any, is attributable to spin-spin correlations.
It has been shown in~\cite{puri1} that the the said conditions are
satisfied by the following criterion:

{\it A quantum state of a system of N spin-1/2s is classical if the
joint quasiprobability for the eigenvalues of the components of each
spin in three mutually orthogonal directions, one of which is the
average direction of that spin, is classical in the symmetric ordering
of the operators. It is non-classical if any of those m-spin
($m\le N$) joint quasiprobabilities is negative in the said ordering.}

Hence, according to this criterion, the JQP for classifying the states
of two spin-1/2s as classical or non-classical is the one given in
(\ref{8p}) where one of the three orthogonal directions ${\bf a}_1,
{\bf b}_1,{\bf c}_1$ is the average direction of spin number 1 and one
of the three orthogonal directions ${\bf a}_2,{\bf b}_2, {\bf c}_2$
is the average direction of spin number 2 with similar interpretation
for the expression in (\ref{8p3c}) for the JQP for three spin-1/2s.

When applied to pure states of $N$ spin-1/2s, the criterion above
identifies~\cite{puri1}, (i) the spin coherent state of $N$
spin-1/2s as classical, (ii) any non-factorizable pure state of two
spin-1/2s as non-classical which is in agreement with the conclusion
in~\cite{gisin} for two spin-1/2s arrived at differently, (iii) any
collective spin state $|N,m\rangle$ (with $\hat S_a|N,m\rangle=
m|N,m\rangle$) as non-classical if $m\ne\pm N/2$, and (iv) a squeezed
state of $N$ spin-1/2s as non-classical for all $N\ge 2$. Thus, except
the states in which all the spins are uncorrelated, all the other pure
states of spin-1/2s examined by applying the said criterion turn out
to be non-classical. This is analogous to the result that, other
than its coherent state, any pure state of the electromagnetic field
is non-classical~\cite{cahil,puri}.

However, when applied to mixed states, it is found that the criterion
above may not identify all the separable states as classical and that
it does not identify those non-separable states as classical which
have been shown to be so by following other methods. Those
deficiencies of the criterion can be remedied by modifying it to read
as follows:

{\it A quantum state of a system of N spin-1/2s is classical 
(i) if joint quasiprobability (JQP) for the eigenvalues of
the components of each spin in three mutually orthogonal directions,
one of which is the average direction of that spin, is classical in
the symmetric ordering of the operators. The state in this case is
also separable.
\noi
or
\noi
(ii) if the JQP for two of the three orthogonal components prescribed
as above is classical. The state in this case may or may not be
separable.}

The states identified as classical as per the condition in (i) in the
criterion above will, of course, be so according to the condition
in part (ii) as well. However, the states identified as classical as
per part (i) are separable which need not be the case for those
identified as classical as per part (ii). It should be emphasized that
the criterion, based as it is on the non-negativity of suitably
defined JQP, is meant basically to identify classical states. The
separability is not related with non-negativity of the JQP. The
separability conjectured in the criterion is based, not on any
fundamental consideration, but on the inference drawn from the study
of separability of some systems.

In the next section we demonstrate the validity of the criterion above
by investigating the classicality and separability of some widely
studied mixed states of two spin-1/2s and a state of three spin-1/2s.

\section{JQP Based Classicality and separability of Some Mixed States}
~\label{sec3}

Let the spin number $j$ be described in the basis constituted by the
eigenstates $|\pm a_j,j\rangle$, corresponding to the eigenvalues
$\pm 1/2$ of its component $\hat S^{(j)}_{a_j}$ in the direction
$\vec{a}_j$. In what follows, we will deal with systems of two and
three spin-1/2s such that  $\vec{a}_1=\vec{a}_2=\vec{a}_3
\equiv\vec{a}$, we denote the basis states of the $j^{{\rm th}}$ spin
simply by $|\pm,j\rangle$. We describe a combined state of the system
in terms of the basis of direct product orthonormal states
$|\pm,\pm\rangle \equiv |\pm,1 \rangle\otimes|\pm,2\rangle$ or their
symmeterised orthonormal combinations,
\be
|\psi_{0,1}\rangle={1\over \sqrt{2}}\left(|+,-\rangle
\mp|-,+\rangle\right),\qquad
|\psi_{2,3}\rangle={1\over\sqrt{2}}\left(|+,+\rangle
\pm|-,-\rangle\right).
\label{7}
\ee
The state $|\psi_0\rangle$, antisymmetric under the exchange of spins,
is the singlet whereas the other three, symmetric under the exchange
of spins, constitute the triplet. Siimilarly, the basis states of
three spin-1/2s shall be denoted by $|\pm,\pm, \pm\rangle \equiv
|\pm,1 \rangle\otimes|\pm,2\rangle\otimes|\pm,3\rangle$.

In the examples below, the average direction of the spins is
identical. The JQPs in (\ref{8p}) and (\ref{8p3c}) then assume the
form
\be
&&p\left(\epsilon^{(1)}_a,\epsilon^{(1)}_b,
\epsilon^{(1)}_c;\epsilon^{(2)}_a,\epsilon^{(2)}_b,
\epsilon^{(2)}_c\right)\nonumber\\
&&=\frac{1}{2^4}{\rm Tr}\left[\hat\rho\prod_{j=1}^{2}\left({1\over 2}
+\epsilon^{(j)}_a \hat S^{(j)}_{a}
+\epsilon^{(j)}_b \hat S^{(j)}_{b}
+\epsilon^{(j)}_c \hat S^{(j)}_{c}\right)\right],
\label{8}
\ee
for the system of two spin-1/2s and, for the system of three
spin-1/2s,
\be
&&p\left(\epsilon^{(1)}_a,\epsilon^{(1)}_b,
\epsilon^{(1)}_c;\epsilon^{(2)}_a,\epsilon^{(2)}_b,
\epsilon^{(2)}_c;\epsilon^{(3)}_a,\epsilon^{(3)}_b,
\epsilon^{(3)}_c\right)\nonumber\\
&&=\frac{1}{2^6}{\rm Tr}\left[\hat\rho\prod_{j=1}^{3}\left({1\over 2}
+\epsilon^{(j)}_a \hat S^{(j)}_{a}
+\epsilon^{(j)}_b \hat S^{(j)}_{b}
+\epsilon^{(j)}_c \hat S^{(j)}_{c}\right)\right],
\label{8c3}
\ee
where $\epsilon^{(j)}_{a,b,c}=\pm 1$; ($\hat S^{(j)}_a,
\hat S^{(j)}_b, \hat S^{(j)}_c$) are the components of the
$j^{{\rm th}}$ spin in the  mutually orthogonal directions
$(\vec{a},\vec{b},\vec{c})$ in which $\vec{a}$ is the
average direction of each of the spins in the state described by
$\hat\rho$. In the examples in which the average value of the spin is
zero, $(\vec{a},\vec{b},\vec{c})$ can be any set of mutually
orthogonal directions. The expectation values involving the
operators $\hat S^{(j)}_{b,c}$ may be evaluated conveniently by noting
that if $\hat S^{(j)}_{a\pm}$ are the raising and lowering operators
on the eigenstates of $\hat S^{(j)}_a$ i.e. if
\be
\hat S^{(j)}_{a+}|-,j\rangle=|+,j\rangle,\qquad
\hat S^{(j)}_{a+}|+,j\rangle=0,\nonumber\\
\hat S^{(j)}_{a-}|+,j\rangle=|-,j\rangle,\qquad
\hat S^{(j)}_{a-}|-,j\rangle=0,
\label{n11}
\ee
then $\hat S^{(j)}_b=(\hat S^{(j)}_{a+}+\hat S^{(j)}_{a-})/2$ and
$\hat S^{(j)}_c=(\hat S^{(j)}_{a+}-\hat S^{(j)}_{a-})/2{\rm i}$.

\vskip .2 in\noi
1. Consider first Werner's density matrix~\cite{werner}
\be
\hat\rho={1-x\over 4}I+x|\psi_0\rangle\langle\psi_0|,
\qquad 0\le x\le 1,
\label{9}
\ee
where $I$ is the identity operator, and $|\psi_0\rangle$ is as in
(\ref{7}). The non-zero expectation values needed for evaluating
the expression in (\ref{8}) in this case are
(with $\langle\hat A\rangle\equiv{\rm Tr}[\hat\rho\hat A]$),
\be
\langle\hat S^{(1)}_\mu\hat S^{(2)}_\mu\rangle=-{x\over 4},
\qquad (\mu=a,b,c).
\label{10}
\ee
The JQP of (\ref{8}) then assumes the form
\be
&&p\left(\epsilon^{(1)}_a,\epsilon^{(1)}_b,
\epsilon^{(1)}_c;\epsilon^{(2)}_a,\epsilon^{(2)}_b,
\epsilon^{(2)}_c\right)\nonumber\\
&&={1\over 2^6}\left[1-x\left(\epsilon^{(1)}_a\epsilon^{(2)}_a
+\epsilon^{(1)}_b\epsilon^{(2)}_b+\epsilon^{(1)}_c\epsilon^{(2)}_c
\right)\right].
\label{11}
\ee
The minimum value of $p$ is
\be
&&p_{{\rm min}}={1\over 2^6}\left[1-3x\right].
\label{12}
\ee
This is non-negative if $x\le 1/3$. Hence, according to the part (i)
of the JQP based criterion, the state (\ref{9}) is classical and also
separable if $x\le 1/3$. This is the same as the condition
for separability of (\ref{9}) according to the Peres
criterion~\cite{peres}. 

\vskip .1 in\noindent
Next, we identify the classical states as per the part (ii) of the
JQP based criterion. To that end,, consider the reduced JQP
$p^{(r)}(\epsilon^{(1)}_a,
\epsilon^{(1)}_b;\epsilon^{(2)}_a,\epsilon^{(2)}_b)$ for the
probability of spin components in the directions ${\bf a}$ and
${\bf b}$ obtained by summing (\ref{11}) over $\epsilon^{(1)}_c$ and
$\epsilon^{(2)}_c$ so that
\be
&&p^{r}\left(\epsilon^{(1)}_a,\epsilon^{(1)}_b;
\epsilon^{(2)}_a,\epsilon^{(2)}_b
\right)\nonumber\\
&&=\frac{1}{2^4}\left[1-x\left(\epsilon^{(1)}_a\epsilon^{(2)}_a
+\epsilon^{(1)}_b\epsilon^{(2)}_b\right)\right].
\label{11n1}
\ee
The minimum value of $p^{r}$ is
\be
&&p^{r}_{{\rm min}}=\frac{1}{2^4}\left[1-2x\right].
\label{12n1}
\ee
This is non-negative if $x\le 1/2$. This shows that the state
(\ref{9}) is classical, not only for the values of $x\le 1/3$ for
which it is separable, but also for the values $1/3<x\le 1/2$ for
which it is not separable. The condition $x\le 1/2$ for the
classicality of the Werner state is the same as the one arrived at
in~\cite{toth} by another method.

\vskip .2 in\noindent
2. Consider the system described by the density matrix~\cite{peres}
\be
\hat\rho= x|\psi_0\rangle\langle\psi_0|+(1-x)|+,+\rangle\langle +,+|,
\qquad 0\le x\le 1.
\label{13n}
\ee
The non-zero expectation values needed for evaluating (\ref{8}) are
\be
&&\langle\hat S^{(1)}_a\rangle=\langle\hat S^{(2)}_a\rangle
={1-x\over 2},\qquad
\langle\hat S^{(1)}_a\hat S^{(2)}_a\rangle={1-2x\over 4},\nonumber\\
&&\langle\hat S^{(1)}_b\hat S^{(2)}_b\rangle
=\langle\hat S^{(1)}_c\hat S^{(2)}_c\rangle=-{x\over 4}.
\label{14n}
\ee
Consequently, the JQP (\ref{8}) is given by
\be
&&p\left(\epsilon^{(1)}_a,\epsilon^{(1)}_b,
\epsilon^{(1)}_c;\epsilon^{(2)}_a,\epsilon^{(2)}_b,
\epsilon^{(2)}_c\right)\nonumber\\
&&=\frac{1}{2^6}\Big[1+(1-x)(\epsilon^{(1)}_a+\epsilon^{(2)}_a)
+\epsilon^{(1)}_a\epsilon^{(2)}_a(1-2x)\nonumber\\
&&-x\left(\epsilon^{(1)}_b\epsilon^{(2)}_b+
\epsilon^{(1)}_c\epsilon^{(2)}_c\right)\Big].
\label{15n}
\ee
The minimum value of the $p$ above may easily be seen to be given
by
\be
&&\left[p\right]_{{\rm min}}=-{x\over 32}.
\label{16n}
\ee
For $0\le x\le 1$, the expression above is non-negative only for
$x=0$ implying that the system admits a LHV description and is
classical only for $x=0$. For this value of $x$, the state (\ref{13n})
is evidently separable. The test of Peres shows that (\ref{13n}) is
indeed separable only for $x=0$. Thus, the JQP based criterion gives
results in agreement with those derived by applying Peres criterion.

\vskip .1 in\noi
3. Consider the density matrix~\cite{horodecki3}
\be
\hat\rho=\sum_{i=0}^{3}x_i|\psi_i\rangle\langle\psi_i|,
\qquad \sum_{i=0}^{3}x_i=1,
\label{27}
\ee
where $|\psi_i\rangle$ are as in (\ref{7}). The non-zero expectation
values in (\ref{8}) in this case are
\be
&&\langle\hat S^{(1)}_a\hat S^{(2)}_a\rangle={1\over 4}(x_2+x_3
-x_0-x_1),\nonumber\\
&&\langle\hat S^{(1)}_b\hat S^{(2)}_b\rangle
={1\over 4}(x_2+x_1-x_0-x_3),\nonumber\\
&&\langle\hat S^{(1)}_c\hat S^{(2)}_c\rangle
={1\over 4}(x_1+x_3-x_0-x_2).
\label{28}
\ee
On substituting these values in (\ref{8}), the JQP corresponding to
(\ref{27}) reads
\be
&&p\left(\epsilon^{(1)}_a,\epsilon^{(1)}_b,
\epsilon^{(1)}_c;\epsilon^{(2)}_a,\epsilon^{(2)}_b,
\epsilon^{(2)}_c\right)\nonumber\\
&&=\frac{1}{2^6}
\Big[1+\epsilon^{(1)}_a\epsilon^{(2)}_a(x_2+x_3-x_0-x_1))
+(\epsilon^{(1)}_b\epsilon^{(2)}_b-\epsilon^{(1)}_c\epsilon^{(2)}_c)
(x_2-x_3)\nonumber\\
&&+(\epsilon^{(1)}_b\epsilon^{(2)}_b+\epsilon^{(1)}_c\epsilon^{(2)}_c)
(x_1-x_0)\Big].
\label{29}
\ee
It may be verified that
\be
&&p(\epsilon_a,\epsilon_b,\epsilon_c;\epsilon_a,\epsilon_b,
\epsilon_c)=\frac{1}{2^5}(1-2x_0),\nonumber\\
&&p(\epsilon_a,\epsilon_b,\epsilon_c;\epsilon_a,-\epsilon_b,
-\epsilon_c)=\frac{1}{2^5}(1-2x_1),\nonumber\\
&&p(\epsilon_a,\epsilon_b,\epsilon_c;-\epsilon_a,-\epsilon_b,
\epsilon_c)=\frac{1}{2^5}(1-2x_2),\nonumber\\
&&p(\epsilon_a,\epsilon_b,\epsilon_c;-\epsilon_a,\epsilon_b,
-\epsilon_c)=\frac{1}{2^5}(1-2x_3).
\label{30}
\ee
The $p's$ for other combinations of the $\epsilon's$ are
non-negative for all values of the $x_i's$. The $p's$ in the
expressions above will also be non-negative i.e. the system will
admit LHV description and will be separable if $x_i\le 1/2$
($i=0,1,2,3$). It may be verified that the same condition is
obtained for the system to be separable according to the Peres
criterion.

\vskip .2 in\noi
4. Consider next the state ~\cite{gisin2}
\be
\hat\rho=x|\psi\rangle\langle\psi|+{1-x\over 2}
\left(|+,+\rangle\langle +,+|+|-,-\rangle\langle -,-|\right),
\label{13}
\ee
where
\be
|\psi\rangle=\alpha|+,-\rangle+\beta|-,+\rangle,\qquad
|\alpha|^2+|\beta|^2=1.
\label{14}
\ee
For the sake of simplicity of illustration, we let $\alpha$ and
$\beta$ to be real. It is then straightforward to show that non-zero
expectation values needed for evaluating (\ref{8}) are given by
\be
&&\langle\hat S^{(1)}_a\rangle=-\langle\hat S^{(2)}_a\rangle
={x\over 2}\left(|\alpha|^2-|\beta|^2\right),
\langle\hat S^{(1)}_a\hat S^{(2)}_a\rangle={1\over 4}(1-2x),
\nonumber\\
&&\langle\hat S^{(1)}_b\hat S^{(2)}_b\rangle
=\langle\hat S^{(1)}_c\hat S^{(2)}_c\rangle=
{x\alpha\beta\over 2}.
\label{15}
\ee
On substituting these values in (\ref{8}), the JQP corresponding to
the state (\ref{13}) reads
\be
&&p\left(\epsilon^{(1)}_a,\epsilon^{(1)}_b,
\epsilon^{(1)}_c;\epsilon^{(2)}_a,\epsilon^{(2)}_b,
\epsilon^{(2)}_c\right)\nonumber\\
&&=\frac{1}{2^6}\Big[1+x(|\alpha|^2-|\beta|^2)(\epsilon^{(1)}_a
-\epsilon^{(2)}_a)
+\epsilon^{(1)}_a\epsilon^{(2)}_a(1-2x)\nonumber\\
&&+2x\alpha\beta\left(\epsilon^{(1)}_b\epsilon^{(2)}_b+
\epsilon^{(1)}_c\epsilon^{(2)}_c\right)\Big].
\label{16}
\ee
It is straightforward to see that the minimum of (\ref{16}) with
respect to $(\epsilon^{(i)}_b,\epsilon^{(i)}_c)$ is achieved when
$\epsilon^{(1)}_b\epsilon^{(2)}_b=\epsilon^{(1)}_c\epsilon^{(2)}_c
=-\alpha\beta/|\alpha||\beta|\equiv s$ so that
\be
&&p\left(\epsilon^{(1)}_a,\epsilon_b,\epsilon_c;
\epsilon^{(2)}_a,s\epsilon_b, s\epsilon_c\right)
\nonumber\\
&&=\frac{1}{2^6}\Big[1+x(|\alpha|^2-|\beta|^2)(\epsilon^{(1)}_a
-\epsilon^{(2)}_a)\nonumber\\
&&+\epsilon^{(1)}_a\epsilon^{(2)}_a(1-2x)-4x|\alpha||\beta|\Big].
\label{17}
\ee
Now, for $\epsilon^{(1)}_a=\epsilon^{(2)}_a=\epsilon$,
\be
&&p\left(\epsilon,\epsilon_b,\epsilon_c;
\epsilon,s\epsilon_b, s\epsilon_c\right)
=\frac{1}{2^5}(1-x-2x|\alpha||\beta|),
\label{18}
\ee
whereas for $\epsilon^{(1)}_a=-\epsilon^{(2)}_a=1$,
\be
&&p\left(1,\epsilon_b,\epsilon_c;
-1,s\epsilon_b, s\epsilon_c\right)
=|\alpha|x(|\alpha|-|\beta|)/16,
\label{19}
\ee
and for $\epsilon^{(1)}_a=-\epsilon^{(2)}_a=-1$,
\be
&&p\left(-1,\epsilon_b,\epsilon_c;
1,s\epsilon_b, s\epsilon_c\right)
=|\beta|x(|\beta|-|\alpha|)/16.
\label{20}
\ee
The JQP's in the equations (\ref{18})-(\ref{20}) are non-negative for
any $x$ if $\alpha\beta=0$. The state (\ref{13}) in this case is
clearly separable. However, if $\alpha\beta\ne 0$ then the JQP's
(\ref{18})-(\ref{20}) are non-negative if $|\alpha|=|\beta|$ and
\be
x\le (1+2|\alpha||\beta|)^{-1}.
\label{21}
\ee
Thus, according to the part (i) of the JQP based criterion, the system
is classical and separable if $|\alpha|=|\beta|$ and $x$ obeys the
condition in (\ref{21}). The condition in the equation above is also
the one for separability according to the Peres criterion~\cite{peres}
but without the additional condition $|\alpha|=|\beta|$. Since
separable states are classical, the set of classical states predicted
by part (i) of the JQP based criterion leaves out  the states having
the value of $x$ as in (\ref{21}) but $|\alpha|\ne |\beta|$. Hence it
is necessary to determine classical states as per part (ii) of that
criterion to find whether the set of those states contains the ones
identified as separable by Peres criterion. To that end, it may be
seen from (\ref{16}) that the JQP for the components along ${\bf b}$
and ${\bf c}$ for each of the two spins is given by
\be
p\left(\epsilon^{(1)}_b,\epsilon^{(1)}_c;\epsilon^{(2)}_b,
\epsilon^{(2)}_c\right)
=\frac{1}{2^4}\Big[1+
2x\alpha\beta\left(\epsilon^{(1)}_b\epsilon^{(2)}_b+
\epsilon^{(1)}_c\epsilon^{(2)}_c\right)\Big].
\label{16n1}
\ee
This is non-negative for all $\alpha,\beta$ if
\be
x\le \frac{1}{4|\alpha||\beta|}.
\label{16n2}
\ee
Keeping in mind the fact that $|\alpha|^2+|\beta|^2=1$, it is seen
that the condition (\ref{16n2}) incorporates (\ref{21}). Hence the JQP
based criterion indeed identifies the set of all separable states as
classical and, in addition, predicts classicality for non-separable
states i.e. the states which satisfy (\ref{16n2}) but not (\ref{21})
as classical. One can construct other two-component JQPs and find
the conditions under which they are non-negative. Those conditions
need not be same as the one found above. However, it is sufficient to
construct one non-negative two-component JQP to identify a state as
classical.

\vskip .2 in\noi
5. Next, we consider a state described by~\cite{horodecki1}
\be
\hat\rho=(1-x)|\phi\rangle\langle\phi|+x|\psi\rangle\langle\psi|,
\label{22}
\ee
where $|\psi\rangle$ is as in (\ref{14}) and
\be
|\phi\rangle=\alpha|+,+\rangle+\beta|-,-\rangle,\qquad
|\alpha|^2+|\beta|^2=1.
\label{23}
\ee
For the sake of simplicity, we let $(\alpha,\beta)$ to be real. The
non-zero expectation values needed for evaluating (\ref{8}) are
given by
\be
&&\langle\hat S^{(1)}_a\rangle={1\over 2}(|\alpha|^2-|\beta|^2),\qquad
\langle\hat S^{(2)}_a\rangle={(1-2x)\over 2}(|\alpha|^2-|\beta|^2),
\nonumber\\
&&\langle\hat S^{(1)}_a\hat S^{(2)}_a\rangle={1\over 4}(1-2x),
\nonumber\\
&&\langle\hat S^{(1)}_b\hat S^{(2)}_b\rangle={\alpha\beta\over 2},
\qquad
\langle\hat S^{(1)}_c\hat S^{(2)}_c\rangle=
{\alpha\beta(1-2x)\over 2}.
\label{24}
\ee
Substitution of these values in (\ref{8}) yields
\be
&&p\left(\epsilon^{(1)}_a,\epsilon^{(1)}_b,
\epsilon^{(1)}_c;\epsilon^{(2)}_a,\epsilon^{(2)}_b,
\epsilon^{(2)}_c\right)\nonumber\\
&&={1\over 64}\Big[1+(|\alpha|^2-|\beta|^2)(\epsilon^{(1)}_a+(1-2x)
\epsilon^{(2)}_a)+\epsilon^{(1)}_a\epsilon^{(2)}_a(1-2x)\nonumber\\
&&+2\alpha\beta\left(\epsilon^{(1)}_b\epsilon^{(2)}_b+(1-2x)
\epsilon^{(1)}_c\epsilon^{(2)}_c\right)\Big].
\label{25}
\ee
With $s=-\alpha\beta/|\alpha||\beta|$, it follows that
\be
&&p\left(1,\epsilon_b,\epsilon_c;
1,s\epsilon_b,s\epsilon_c\right)
=(1-x)|\alpha|(|\alpha|-|\beta|)/16,\nonumber\\
&&p\left(1,\epsilon_b,\epsilon_c;
1,s\epsilon_b,-s\epsilon_c\right)=
|\alpha|[(1-x)|\alpha|-|\beta|x]/16,\nonumber\\
&&p\left(1,\epsilon_b,\epsilon_c;
-1,s\epsilon_b,-s\epsilon_c\right)=
|\alpha|x(|\alpha|-|\beta|)/16,\nonumber\\
&&p\left(1,\epsilon_b,\epsilon_c;
-1,s\epsilon_b,s\epsilon_c\right)=
|\alpha|[|\alpha|x-(1-x)|\beta|]/16,\nonumber\\
&&p\left(-1,\epsilon_b,\epsilon_c;
1,s\epsilon_b,s\epsilon_c\right)=
|\beta|[x\beta-(1-x)|\alpha|]/16,\nonumber\\
&&p\left(-1,\epsilon_b,\epsilon_c;
-1,s\epsilon_b,-s\epsilon_c\right)
=|\beta|x(|\beta|-|\alpha|)/16,\nonumber\\
&&p\left(-1,\epsilon_b,\epsilon_c;
-1,s\epsilon_b,s\epsilon_c\right)=
|\beta|(1-x)(|\beta|-|\alpha|)/16,\nonumber\\
&&p\left(-1,\epsilon^{(1)}_b,\epsilon^{(1)}_c;
-1,s\epsilon_b,-s\epsilon_c\right)
=|\beta|[(1-x)|\beta|-x|\alpha|]/16.
\label{26}
\ee
The JQP (\ref{25}) for combinations of the $\epsilon's$ other than
those appearing in the equations above are non-negative for all values
of $(x,\alpha,\beta)$. The JQPs in the equation above will also be
non-negative for all $0\le x\le 1$ if $\alpha\beta=0$. In this case
the density matrix (\ref{22}) is clearly separable.

If $\alpha\beta\ne 0$ then the $p's$ in (\ref{26}) are non-negative if
$|\alpha|=|\beta|$ and $x=1/2$ in which case the given state is
classical and separable as per the part (i) of the JQP based
criterion. However, the separability criterion of Peres
leads to the condition $x=1/2$~\cite{horodecki1} without the
restriction $|\alpha|=|\beta|$. Since separable states are classical,
we look for the left out classical states by the part (i) of the
criterion by examining the JQP for two orthogonal components as per
its part (ii). To that end, note from (\ref{25}) that the JQP for the
components in the directions ${\bf b}$ and ${\bf c}$ for spin 1 and
those in the directions ${\bf a}$ and ${\bf c}$ for spin 2 is given by
\be
p\left(\epsilon^{(1)}_b,\epsilon^{(1)}_c;\epsilon^{(2)}_a,
\epsilon^{(2)}_c\right)
={1\over 16}\left[1+(1-2x)\{(|\alpha|^2-|\beta|^2)\epsilon^{(2)}_a
+2\alpha\beta\epsilon^{(1)}_c\epsilon^{(2)}_c\}\right].
\label{25n1}
\ee
This is non-negative if $x=1/2$ for all $\alpha,\beta$. The JQP based
criterion thus identifies all those states as classical which are
found to be separable according to Peres criterion. As stated
before, one may find additional classical states by examining the
positivity of the JQPs corresponding to other combinations of two of
the three components.

\vskip .2 in\noi
6. Lastly, we show that the condition for classicality predicted by
the JQP based criterion of the following state of a system of three
spin-1/2s, introduced in~\cite{toth}, is the same as the one obtained
in that reference by another method:
\be
\hat\rho&=&\frac{I}{8}+\frac{1}{6}\sum_{\mu=x,y,z}
\hat S^{(2)}_\mu\hat S^{(3)}_\mu
-\frac{c}{4}\sum_{\mu=x,y,z}\left[\hat S^{(1)}_\mu
\hat S^{(3)}_\mu
+\hat S^{(1)}_\mu\hat S^{(2)}_\mu\right].
\label{3spin1}
\ee
Note that $\langle\hat S^{(i)}_\mu\rangle=0$. We choose the
$z$-direction as the average direction ${\bf a}$ of the spins. Recall
the expression (\ref{8c3}) for the JQP for three spins and evaluate
required averages to get
\be
&&p\left(\epsilon^{(1)}_a,\epsilon^{(1)}_b,\epsilon^{(1)}_c;
\epsilon^{(2)}_a,\epsilon^{(2)}_b,\epsilon^{(2)}_c;
\epsilon^{(3)}_a,\epsilon^{(3)}_b,\epsilon^{(3)}_c\right)\nonumber\\
&&=\frac{1}{2^9}\Big[1+\frac{1}{3}\left\{
\epsilon^{(2)}_a\epsilon^{(3)}_a+\epsilon^{(2)}_b\epsilon^{(3)}_b
+\epsilon^{(2)}_c\epsilon^{(3)}_c\right\}\nonumber\\
&&-\frac{c}{2}\left\{\epsilon^{(1)}_a\left(
\epsilon^{(2)}_a+\epsilon^{(3)}_a\right)
+\epsilon^{(1)}_b\left(\epsilon^{(2)}_b+\epsilon^{(3)}_b\right)
+\epsilon^{(1)}_c\left(\epsilon^{(2)}_c+\epsilon^{(3)}_c\right)
\right\}\Big].
\label{3spin2}
\ee
It is straightforward to verify that $p$ above is non-negative if
$c\le 2/3$. This shows that, according to the part (i) of the JQP
based criterion, the state is classical and separable if
$c\le 2/3$.

\vskip .1 in\noi
To find the other set of classical states, we invoke the part (ii)
of the criterion. To that end, we construct from (\ref{3spin2}) the
reduced distribution for the directions ${\bf b},{\bf c}$ for spin 1
and the directions ${\bf a},{\bf b}$ for spins 2 and 3 to get
\be
&&p^{(r)}\left(\epsilon^{(1)}_b,\epsilon^{(1)}_c;
\epsilon^{(2)}_a,\epsilon^{(2)}_b;
\epsilon^{(3)}_a,\epsilon^{(3)}_b\right)\nonumber\\
&&=\frac{1}{2^6}\Big[1+\frac{1}{3}\left\{
\epsilon^{(2)}_a\epsilon^{(3)}_a+\epsilon^{(2)}_b\epsilon^{(3)}_b
\right\}-\frac{c}{2}\left\{
\epsilon^{(1)}_b\left(\epsilon^{(2)}_b+\epsilon^{(3)}_b\right)
\right\}\Big].
\label{3spin4}
\ee
The probability above is non-negative if $c\le 1$. It may be verified
that the upper bound on the value of $c$ for which the JQPs
corresponding to other combinations of two of the three components are
non-negative is less than 1. Hence the JQP based criterion predicts
classicality of the three spins-1/2 state (\ref{3spin1}) if $c\le 1$.
This is in agreement with the result of~\cite{toth} that $\hat\rho$ in
(\ref{3spin1}) admits LHV description if $c\le 1$.

It is also of interest to study the properties of the reduced density
matrix corresponding to the spins numbered 1 and 2. It is obtained by
summing (\ref{3spin2}) over $\epsilon^{(3)}_\mu$ ($\mu=a,b,c)$ and reads
\be
p\left(\epsilon^{(1)}_a,\epsilon^{(1)}_b,\epsilon^{(1)}_c;
\epsilon^{(2)}_a,\epsilon^{(2)}_b,\epsilon^{(2)}_c\right)
=\frac{1}{2^6}\Big[1-\frac{c}{2}
\left(\epsilon^{(1)}_a\epsilon^{(2)}_a
+\epsilon^{(1)}_b\epsilon^{(2)}_b
+\epsilon^{(1)}_c\epsilon^{(2)}_c\right)\Big].
\label{3spin3}
\ee
The probability in the expression above is evidently positive if
$c\le 2/3$. This implies that the reduced density matrix is classical
and separable for $c\le 2/3$. This is in agreement with the result
reported in~\cite{toth} that the reduced density matrix corresponding
to spins 1 and 2 exhibits entanglement if $c>2/3$. The same results
hold for the system of spins 1 and 3 described by the corresponding
reduced density matrix.

\section{Conclusions}~\label{sec4}

Based on the concept of joint quasiprobability distribution of the
eigenvalues of the components of spins, a criterion has been proposed
to identify the classical states of a system of spin-1/2s. Its
validity has been demonstrated by applying it to mixed states of such
systems of two spin-1/2s and a system of three spin-1/2s whose
classicality and separability properties are known by other methods
and showing that the proposed criterion gives results in agreement
with the known ones. The criterion offers a unified approach to study
classicality of a system of any number of spin-1/2s and also
offers the possibility of identifying the processes of
measurement~\cite{pop1}-\cite{horodeckin}
under which the predicted classicality can be observed.

\vskip .1 in\noi
Acknowledgement: The author would like to thank one of the referees
for very useful comments.

\end{document}